\documentclass{PoS}
\usepackage{wrapfig}
\usepackage{cite}
\def\vector#1{\mbox{\boldmath ${#1}$}} %
\def\v#1{\vector{#1}} 
\def\m#1{\mathrm{#1}} 
\def\tendto{\rightarrow} 
\def\be#1{\begin{equation}#1\end{equation}} 

\def\beqn#1{\begin{eqnarray*}#1\end{eqnarray*}}
\newcommand{\diff}{\mathrm{d}}  
\def\pdif#1#2{\frac{\partial {#1}}{\partial {#2}}} 
\def\exp#1{\mathrm{e}^{#1}} 
\def\bra#1{\langle{#1}|} 
\def\cket#1{|{#1}\rangle} 

\def\integral#1#2{\int^{#2}_{#1}} 
\def\summation#1#2{\sum^{#2}_{#1}} 

\title{Charmonium current-current correlators with Mobius domain-wall fermion}

\ShortTitle{Charmonium current-current correlators}

 \author{\speaker{Katsumasa Nakayama}$^{a, b}$, Brendan Fahy$^b$, and Shoji Hashimoto$^{b,c}$ (JLQCD collaboration)\\
         $^a$ Department of Physics, Nagoya University, Nagoya, 464-8602, Japan\\
         $^b$ KEK Theory Center, High Energy Accelerator Research Organization (KEK), Tsukuba 305-0801, Japan\\
         $^c$ School of High Energy Accelerator Science, The Graduate University for Advanced Studies (Sokendai),Tsukuba 305-0801, Japan\\
        E-mail: \email{katumasa@post.kek.jp}}

\abstract{
We calculate the charmonium correlators on the lattice with $n_f = 2+ 1$ Moebius domain wall fermion, and extract the charm quark mass 
and the strong coupling constant. 
Time moments are defined by current-current correlators, which have been calculated in the continuum theory by perturbation theory. 
We extract the charm quark mass by matching the lattice results with the corresponding perturbative QCD calculations, using the recently generated ensembles by the JLQCD collaboration at lattice spacings $a$ = 0.083, 0.055, and 0.044 fm.
}

\FullConference{The 33rd International Symposium on Lattice Field Theory\\
		14 -18 July 2015\\
		Kobe International Conference Center, Kobe, Japan*}

\begin{document}

\section{Introduction}

The charm quark mass $m_c$ and the strong coupling constant $\alpha _s$ are fundamental parameters in the Standard Model. 
Their precise determination is important for the test of the Standard Model.
One example is related to Higgs partial widths, which has significant dependence on $\alpha _s$ and $m_c$. For the determination of the partial widths better than 1\%, one needs $m_c$ and $\alpha _s$ also better than 1\% 
\cite{HiggsBR2}.

We extract the charm quark mass and the strong coupling constant using the moment method in lattice QCD with the Mobius domain-wall fermion.
We calculate the charmonium current-current correlator on the lattice and construct its time moments, which correspond to derivatives of the vacuum polarization functions.
The moments can be related to the R-ratio $R(e^+e^- \tendto q\bar{q} +X)$ by the  dispertion relation, and one can obtain the charm quark mass using experimental data and perturbation theory. Instead of the experimental data, we use the lattice data to extract the charm quark mass. The idea was initiated by the HPQCD collaboration, and a precision better than 1\% has been reported \cite{HPQCD2007,HPQCD2010,HPQCD2015}.

\section{Moment method}
%
%
We calculate the charmonium pseudoscalar correlator summed over spatial position $\v{x}$
\be{
G(t) = a^6 \summation{\v{x}}{} (am^{\m{bare}} _\m{quark})^2 \bra{0} j_5 (\v{x},t)j_5 (0,0)\cket{0},
}
with $j_5 = \bar{\psi _c}\gamma _5 \psi _c$. The moment of the pseudoscalor current-current correlator on the lattice is defined by
\be{
G_n = \summation{t}{}\left( \frac{t}{a}\right) ^n G(t),
\label{momentdef}
} 
with $n$ an even integer larger than 4. On the lattice $t/a$ takes a value in $\{ 0, 1, 2,...,\frac{T}{2a} - 1, 0, -\frac{T}{2a} + 1,...,-2,-1\} ,$
where $T$ is the size of the lattice in the time direction. Because of the symmetry $G(t) = G(- t)$ odd $n$ moments vanish.

We follow the method introduced in \cite{HPQCD2007,HPQCD2010}. To reduce the discretization effect, we define the reduced moment $R_n$ using the lowest order moment $G_n ^{(0)}$ evaluated for a free correlator:
\be{
R_n = \frac{am_{\eta_c} ^{\m{lat}}}{2am_{\m{bare,\ c}} ^{\m{lat}}}\left(\frac{G_n}{G_n ^{(0)}}\right)^{\frac{1}{n-4}}
}
 where $m_{\eta_c} ^{\m{lat}}$ is the $\eta _c$ mass on each lattice ensemble, and $m_{\m{bare,\ c}} ^{\m{lat}}$ is the bare charm quark mass on that ensemble.

The moments in the continuum perturbation theory are similarly defined by the correlators of $j_5(x)$. First, we consider the pseudoscalor vacuum polarization function $\Pi (q^2)$,
\be{
q^2 \Pi (q^2) = i \integral{}{} \diff x \exp{iqx} \bra{0} T j_5(x)j_5(0)\cket{0}.
}
By a Taylor expansion, it may be expressed as
\be{
\Pi (q^2) 
= \frac{3}{16\pi ^2} Q_q ^2\summation{k = -1}{\infty} C_k z^k,
\label{expanvpf}
}
where $z$ = $\left( {q}/{2m_\m{c}}\right) ^2$, $Q_q$ is the quark charge (+2/3 or -1/3), and $C_k$ are coefficients calculated perturbatively up to $O(\alpha _s ^3)$ \cite{HPQCD2010, Fourloopmoment1,Fourloopmoment2, Fourloopmoment3}. They are written as
\beqn{
C_k &=& C_k ^{(0)} + \frac{\alpha _s (\mu)}{\pi} \left( C_k ^{(10)} + C_k ^{(11)}l_{m}\right) \\
& &\ \ \ + \left( \frac{\alpha _s (\mu)}{\pi}\right) ^2 \left( C_k ^{(20)} + C_k ^{(21)}l_{m} + C_k ^{(22)}l_{m} ^2 \right) \\
& &\ \ \ + \left( \frac{\alpha _s (\mu)}{\pi}\right) ^3 \left( C_k ^{(30)} + C_k ^{(31)}l_{m} + C_k ^{(32)}l_{m} ^2 + C_k ^{(33)}l_{m} ^3 \right) \\
& &\ \ \ +...
}
with $l_m = \log{(m^2 _c (\mu)/\mu ^2)}$.

Since the $t$ multiplication on the lattice corresponds to a differentiation by $q^2$ in the momentum space, the moment may be written as
\be{
g_{2n}  = \left( \frac{1}{n!}\right)\left( -\pdif{}{z}\right) ^n (z\Pi (q^2))|_{q^2 = 0},
}
and one can calculate the reduced moment in the continuum theory as $r_{2k} =({g_{2k}}/{g_{2k} ^{(0)}})^{{1}/{(2k-4)}}= ({C_{k-1}}/{C_{k-1} ^{(0)}}) ^{{1}/{(2k-4)}}$,
 where $C^{(0)} _{k-1}$ are the coefficients at the lowest order.
Requiring the equality between the lattice and the continuum, we arrive at
\be{
m_c (\mu ) =\frac{m^{\m{exp}} _{\eta_c}}{2}\frac{r_n (\alpha _{\m{\overline{MS}}},m_c)}{R_n},
\label{massextra}
}
where $m_c(\mu)$ is the $\overline{\mathrm{MS}}$ charm quark mass defined at a renormalization scale $\mu$. $r_n$ appearing on the right hand side is an implicit function of $m_c(\mu)$ and $\alpha _s (\mu)$, and the equation should be understood as a condition to be satisfied by the parameters $m_c(\mu)$ and $\alpha _s (\mu)$. Using the equation for different $n$'s at the same time, we may determine these parameters. 
We may also use a ratio of the reduced moments, 
\be{
\frac{R_n }{R_{n+2} } = \frac{r_n }{r_{n+2} },
}
for which the truncation error of the perturbative expansion is different from that of individual $r_n$.
\begin{figure}[htbp]
\begin{center}
\includegraphics[width=8.0cm, angle=0]{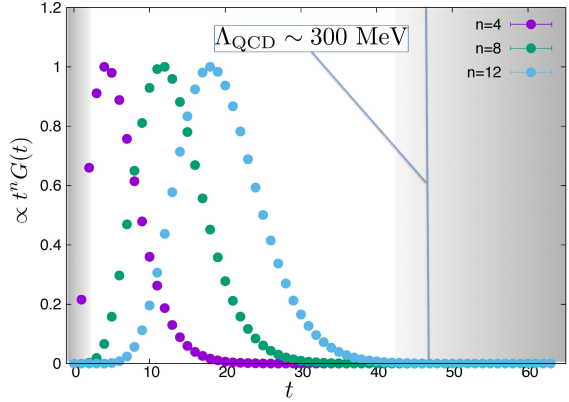}
\caption{$t^n G(t)$ on the lattice size $64^3\times 128$ at $a=0.044$ fm. The peak is normalized to 1. The data for $n=4$, 8, and 12 are shown.}
\label{window}
\end{center}
\end{figure}

The moment method has a range of validity for the scale $\mu$ since we concurrently use the lattice and perturbation theory.
That is, there is a upper bound on $n$ from $\Lambda _{\m{QCD}}$ for perturbativation theory to be valid. On the other hand, the lower bound comes from the lattice cutoff. 
Figure \ref{window}
 demonstrate the range of $t$ that $R_n$ recieves dominant contribution. The integrand of (\ref{momentdef}) is plotted as a function of $t$, for $n$=4, 8, and 12. The peak of the integrand is estimated as $t_{\m{peak}} \sim {n}/{m_{\eta_c} ^{\m{lat}}}$, and the valid range ${\pi}/{a} \gg \pi/t_{\m{peak}}\gg \Lambda _{\m{QCD}}$, is interpreted as
\be{
m_{\eta_c}  ^{\m{lat}} a \ll n \ll \pi\frac{m_{\eta_c}  ^{\m{lat}}}{\Lambda _{\m{QCD}}}.
\label{scale}
}
As one can see in 
Fig. \ref{window},
 $R_4$ receives a large contribution from small $t$ ($t = 1, 2$), which may have relatively large discretization effect. As the power $n$ becomes larger, the moment receives more contributions from large $t$ region. Assuming $\Lambda _{\m{QCD}} \sim 300$ MeV, the constraint (\ref{scale}) corresponds to 
$n \ll 30$. We therefore use $n = 6,8,10$ in this analysis.
\section{Analysis and error estimation}

Our lattice QCD simulations are carried out with $n_f=2+1$ Moebius domain wall fermion at lattice spacings $a=0.083$, 0.055, and 0.044 fm. The spatial size of the lattice is $L$ = 32, 48, and 64 depending on $a$ and the temporal size $T$ is twice the spatial size $L$. For the details of the lattice ensembles, see \cite{brendan}. On each ensemble, we calculate the charmonium correlator at three different bare charm quark masses.

We calculate the reduced moment $R_n$ on each ensemble, interpolate them to the physical charm quark mass, and then extrapolate to the continuum and the chiral limit of light quarks. 
First we interpolate in $m_c$ to the physical point by tuning until the spin average mass $({m_{\eta_c} + 3m_{J/\psi}})/{4}$ 
reproduces the experimental value obtained from the PDG value ($m_{\eta_c} ^{\m{exp}} = 2.9836(7)$ GeV and $m_{J/\psi} ^{\m{exp}} = 3.0969$ GeV)
. We then extrapolate $R_n$ assuming the form
\be{
R_n = R_n(0) \left( 1 + c_1(am_c) ^2 \right) \times \left( 1 + f_1\frac{m_u + m_d + m_s}{m_c} \right),
\label{fittingfunc}
}
with free parameters $R_n(0)$, $c_1$, and $f_1$.
The error of $O(a^2)$ is eliminated by an extrapolation with this form, while the effect of $O(a^4)$ still needs to be estimated. We take the value of $c_1$ from this fit as a typical size of the coefficients also at higher orders, and artificially add or subtract $c_1 (am_c)^4$ to the term representing the discretization effect ($1+c_1(am_c)^2$). Namely, we modify the fit form as 
\be{
R_n = R ^{(\pm)} _n(0) \left( 1 + c ^{(\pm)}_1(am_c) ^2 \pm c ^{(0)}_1(am_c) ^4 \right) \times \left( 1 + f ^{(\pm)}_1\frac{m_u + m_d + m_s}{m_c} \right),
}
%
with $c_1^{(0)}$ fixed from (\ref{fittingfunc}) while other parameters are free. We repeat the fitting and take the largest variation of $R_n(0) ^{(\pm)}$ as an estimate of the remaining discretization error.

The quark mass dependence, which is assumed to be linear in $m_u + m_d + m_s$, turned out to be flat ($f_1 \sim 0$), 
and we do not consider its higher order effects. We can also neglect the effect of small non-zero values of $m_u + m_d + m_s$ at the physical point.
Table \ref{RRR} is the results for $R ^{(0)}_n$. We use the standard $\chi ^2$ fitting and the statistical error is estimated through the covariance matrix. Finite volume error is estimated by inspecting the difference between the results at $L=32$ and at 48 on the coarsest lattice.
\begin{table}
\begin{minipage}{.45\textwidth}
\begin{center}
\begin{tabular}{|c||c|} \hline
&(Stat.)($a$)(O$(a^4)$)(Vol.) \\ \hline
$R_6 ^{(0)}$&1.520(2)(1)(8)(5)		 \\ \hline
$R_8 ^{(0)}$&1.368(1)(1)(4)(2)	 \\ \hline
$R_{10} ^{(0)}$&1.302(1)(0)(3)(1)	 \\ \hline
$R_{12} ^{(0)}$&1.262(1)(0)(3)(0)		 \\ \hline
$R_{14} ^{(0)}$&1.236(1)(0)(3)(0)	 \\ \hline
\end{tabular}
\label{Rn}
\end{center}
\end{minipage}
\begin{minipage}{.45\textwidth}
\begin{center}
\begin{tabular}{|c||c|} \hline
&(Stat.)($a$)(O$(a^4)$)(Vol.) \\ \hline
$R_6  ^{(0)}/R_8  ^{(0)}$&1.1113(6)(3)(28)(2)	 \\ \hline
$R_8  ^{(0)}/R_{10}  ^{(0)}$&1.0510(2)(1)(5)(1)	 \\ \hline
$R_{10}  ^{(0)}/R_{12}  ^{(0)}$&1.0313(1)(1)(0)(1)	 \\ \hline
\end{tabular}
\label{ratioRn}
\end{center}
\end{minipage}
\caption{Reduced moment in the continuum limit (left) and their ratio (right). The error from statistical, the input of the lattice spacing, $O(a^4)$, and finite volume effect are also listed.}
\label{RRR}
\end{table}
%
%
%
%
%
%
%
%
%
%
%

\begin{figure}[t]
\begin{center}
\includegraphics[width=7.5cm, angle=0]{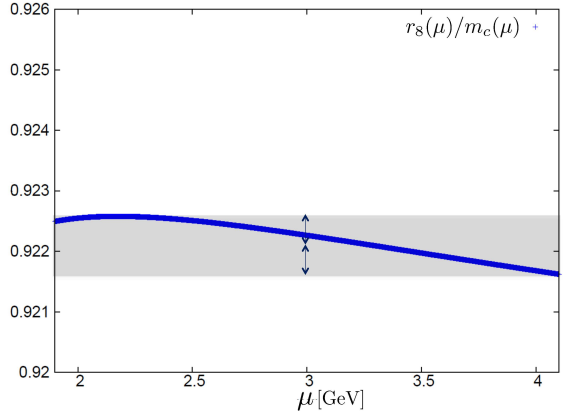}
\caption{$\mu$ dependence $r_8(\mu)/m_c(\mu)$ for $n=8$}
\label{mu8depend}
\end{center}
%
%
\end{figure}

For the value of $m_{\eta_c} ^{\m{exp}}$ in (\ref{massextra}) we input the experimental value 2.9836(7) GeV after subtracting the corrections due to disconnected and electromagnetic effects.
Disconected effect is estimated as $m_{\eta_c} - m_{\eta_c} ^{(\m{no\ disconect})} = -2.4(8)\m{\ MeV}$ \cite{disconected2}, while the electromagnetic contribution is $m_{\eta_c} - m_{\eta_c} ^{(\m{no\ EM})} = -2.6(1.3)\m{\ MeV}$ in \cite{EM}. 
The hyperfine splitting $\Delta _{J/\psi - \eta_c}=m_{J/\psi} - m_{\eta_c}$ is directly calculated on the lattice. Our lattice data fail to obtain the hyperfine splitting consistent with the experimental data 113.3 MeV at finite lattice spacings. After extrapolating to the continuum limit we obtain 115.7(17) MeV. We estimate the error due to higher order effect of $O(a^4)$ by adopting different extrapolations, and the associated error for $m_{\eta_c}$ is estimated as $2.3$ MeV.
%
%
Including all the errors in the $\eta _c$ meson mass, the input is $
m_{\eta_c} ^{\m{exp}}= 2983.6(0.7) + 2.4(0.8)_{\m{Disc.}} + 2.6(1.3)_{\m{EM}}  \pm (2.3)_{\m{HF}} \m{\ MeV}$.

Perturbative calculation is available up to $O(\alpha _s ^3)$, and remaining error is of $O(\alpha _s ^4)$. We estimate the truncation error from the residual $\mu$ dependence of ${r_n(\mu)}/{m_c(\mu)}$. We take $\mu = 3$ GeV as a central value and consider the variation in the range $\pm 1$ GeV for the estimate of the truncation error.
 Figure \ref{mu8depend}
 shows an example for $n = 8$. The $\mu$ dependence of $r_n(\mu)$ is almost canceled by the dependence of $m_c(\mu)$, and the remnant $\mu$ dependence is tiny but non-zero which we take as the truncation error. Such cancellation also occurs for the ratio of the reduced moment.
 %
\begin{figure}[t]
\begin{center}
\includegraphics[width=9.0cm, angle=0]{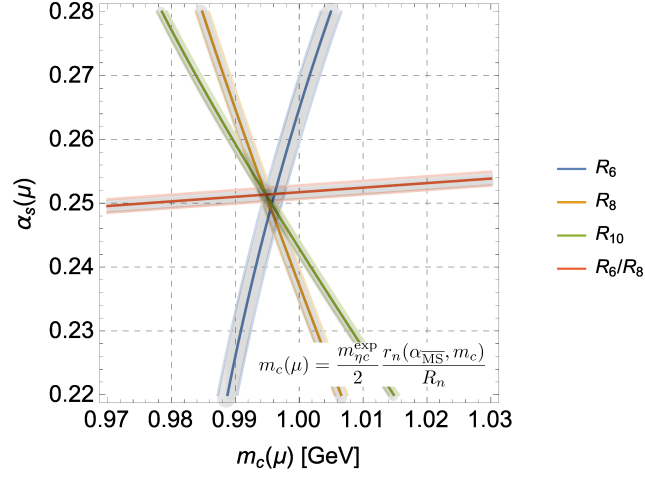}
\caption{\small The solution of $R_6/R_8$, $R_8$, $R_{10}$ on the $\{m_c(\mu),\ \alpha _s(\mu)\}$ plane. The bands express the statistical error.}
\label{solgra}
\end{center}
\end{figure}

Next, we consider the non-perturbative correction on the perturbative side. 
The perturbative expansion is associated with the non-perturbative correction represented by power corrections in the operator product expansion. Namely, we replace $C_{k-1}$ in (\ref{expanvpf}) by
\be{
C_{k-1}\tendto C_{k-1} + \frac{16\pi ^2}{3Q_q ^2}\frac{\langle(\alpha _s /\pi )G^{\mu\nu}G_{\mu\nu}\rangle}{(2m_c)^4}A_k,
}
where the constants $A_k$ are given in \cite{gluecond}. 
The expectation value $\langle(\alpha _s /\pi )G^{\mu\nu}G_{\mu\nu}\rangle$, is called the gluon condensate, and its precise value is not known. We therefore keep it a free parameter and determine by a fit of $R_n$.

Over all, we extract the charm quark mass $m_c(\mu)$, the strong coupling constant $\alpha _s(\mu)$, and the gluon condensate ${\langle(\alpha _s / \pi) G^2\rangle}/{m _c ^4}$, simultaneously. 
Figure \ref{solgra} shows the constraint from $r_n$ on $m_c(\mu)$ and $\alpha _s (\mu)$ for $n=$6, 8, and 10 as well as that from the ratio $R_6/R_8$. We observe that the bands from each $r_n$ cross at the same point in the ($m_c$, $\alpha _s$) plane after adjusting $\langle(\alpha _s /\pi )G^{\mu\nu}G_{\mu\nu}\rangle$ appropriately.

%
%
We use three constraints at once to obtain the results for $m_c(\mu)$, $\alpha _s(\mu)$, and ${\langle(\alpha _s / \pi) G^2\rangle}/{m _c ^4}$. The best choice among the different possibilities turned out to be a combination of $R_6/R_8, R_8,$ and $R_{10}$.
The result is 
%
\be{
m_c(\mu = 3\m{\ GeV})=0.9948(26)(16)(64)\m{\ GeV}, 
}
\be{
\alpha _{\overline{\m{MS}}}(\mu = 3\m{\ GeV})=0.2514(74)(11)(58)\m{\ GeV},
}
and ${\langle (\alpha _s /\pi )G^{\mu\nu}G_{\mu\nu}\rangle}/{m^4 _c}=0.0007(14)(0)(1)$, 
where the errors represent the perturbative truncation error, statistical errors, and other systematic errors, respectively. Statistical errors contain those of the configurations and from the error of the lattice spacing. Systematic uncertainty is from finite volume effect, $O(a^4)$ error, disconnected and electromagnetic contributions, hyperfine splitting, and input experimental error in the $m_{\eta_c} ^{\m{exp}}$. Each contribution is summarized in Table \ref{typexam}, for $m_c(3\m{\ GeV})$ and $\alpha _s (3\m{\ GeV})$.
\begin{table}
\begin{center}
\begin{tabular}{|c||c|c|} \hline
$R_6/R_8, R_8,$ and $R_{10}$&$m_c(3\m{\ GeV})$	&$\alpha _s (3\m{\ GeV})$				    \\ \hline
Perturbation theory&0.3&2.9			    \\ \hline
Statistical erorrs&0.1&0.4			    \\ \hline
Lattice spacings&0.1&0.2			    \\ \hline
$O(a^4)$&0.5&1.9			    \\ \hline
Finite volume effect&0.3&1.3			    \\ \hline
Input $m_{\eta_c} ^{\m{exp}}$&0.0&0.0			    \\ \hline
Disconnected&0.0&0.0			    \\ \hline
Electromagnetic&0.1&0.0			    \\ \hline
$\eta _c - J/\psi$ hyperfine splitting&0.1&0.0			    \\ \hline
Total&0.7\%&3.8\%			    \\ \hline
\end{tabular}
\caption{Each error contribution to the result from $R_6/R_8, R_8,$ and $R_{10}$.}
\label{typexam}
\end{center}
\end{table}

These results may be converted to those at the scale $\mu = m_c$. We obtain the $m_c(\mu = m_c) = 1.2769(91)$ GeV. Also, for $\alpha _{\overline{\m{MS}}}(\mu)$ at the weak scale, our result is $\alpha _{\overline{\m{MS}}}(\mu = M_Z) = 0.1174(20)$.\\ \\
\textbf{Acknowledgements}

The lattice QCD simulation has been performed on Blue Gene/Q supercomputer at the High Energy Accelerator Research Organization (KEK) under the Large Scale Simulation Program (Nos. 13/14-4, 14/15-10). This work is supported in part by the Grant-in-Aid of the Japanese Ministry of Education (No. 26247043) and the SPIRE (Strategic Program for Innovative Research) Field 5 Project.
%
%
%
%
%
%
%

\end{document}